\documentclass[pra,twocolumn,showpacs]{revtex4}
\usepackage{graphicx}
\usepackage{graphics}
\usepackage{amsfonts}
\usepackage{amsmath}
\usepackage{epsfig}

\begin{document}
\catcode`\ä = \active \catcode`\ö = \active \catcode`\ü = \active
\catcode`\Ä = \active \catcode`\Ö = \active \catcode`\Ü = \active
\catcode`\ß = \active \catcode`\é = \active \catcode`\è = \active
\catcode`\ë = \active \catcode`\ô = \active \catcode`\ê = \active
\catcode`\ø = \active \catcode`\ò = \active \catcode`\í = \active
\catcode`\Ó = \active \catcode`\ú = \active \catcode`\á = \active
\catcode`\ã = \active
\defä{\"a} \defö{\"o} \defü{\"u} \defÄ{\"A} \defÖ{\"O} \defÜ{\"U} \defß{\ss} \defé{\'{e}}
\defè{\`{e}} \defë{\"{e}} \defô{\^{o}} \defê{\^{e}} \defø{\o} \defò{\`{o}} \defí{\'{i}}
\defÓ{\'{O}} \defú{\'{u}} \defá{\'{a}} \defã{\~{a}}

\newcommand{\li}{$^6$Li}
\newcommand{\na}{$^{23}$Na}

\title{Multiple species atom source for laser-cooling experiments}

\author{C. A. Stan and W. Ketterle}

\affiliation{Department of Physics\mbox{,} MIT-Harvard Center for
Ultracold Atoms\mbox{,} and Research Laboratory of Electronics,\\
MIT, Cambridge, Massachusetts 02139}

\date{June 2, 2005}

\begin{abstract}

We describe the design of a single beam, multiple species atom
source in which the flux of any component can be separately
adjusted. Using this design we have developed a \na\ -\li\ atom
source for ultracold atom experiments. The fluxes of lithium and
sodium are independently tunable, allowing operation as a single
\na\ or \li\ source as well as a double source with equal atomic
fluxes in each component.

\end{abstract}

\pacs{03.75.F, 07.77.Gx, 32.80.Pj, 39.10.+j}

\maketitle

\section{Introduction}

In recent years a number of frontier atomic physics experiments have
used more than one atomic species. The most precise measurement of
the electrical dipole moment of the electron to date \cite{rega02}
used two atoms to suppress systematic errors. Ultracold mixtures of
two atomic species are used for sympathethic cooling of fermions
\cite{trus01,schr01,hadz02}, for the study of atomic Bose-Fermi
mixtures \cite{modu02}, and of interspecies Feshbach resonances
\cite{stan04,inou04}. Simultaneous laser cooling of the atomic
components is used to study interspecies collisions
\cite{sant96,lund04}, and to produce ultracold heteronuclear
molecules \cite{shaf99} which might be used for quantum computation
\cite{demi02} or the observation of exotic many-body states
\cite{gora02,bara02}.

Vapor cells and atomic beams are the most common atom sources for
cold atom experiments. In a vapor cell, an atomic trap is loaded
from background gas. Higher background gas pressures results in
faster loading rates, but above an optimum pressure losses due to
collisions with the background gas increase as fast as the loading
rate. Multispecies vapor cells have been used to generate a slow
beam of rubidium and cesium atoms \cite{lund04}.

Loading an atomic trap from a collimated beam of atoms is
compatible with ultrahigh vacuum and avoids background gas
collisions. The flux of atoms slow enough to be captured by the
trap can be considerably increased compared to a thermal beam by
using a Zeeman slower \cite{phil82}. The Zeeman slower system has
higher loading rates than a vapor cell.

A multiple-species experiment can be designed with multiple
independent beam sources, but this increases its complexity. We have
developed a two-species atomic oven for \na\ and \li\ which works
with a single Zeeman slower to load \na\ and \li\ magneto-optical
traps (MOTs) simultaneously.

The paper is organized as follows: in Sec. II we present the general
design and operation criteria for a multiple species oven, and in
Sec. III the construction of the sodium-lithium source. We describe
the operation of this source and its performance in Sec. IV, and we
discuss in Sec. V the applicability of the sodium-lithium design to
other atomic species used in laser-cooling experiments.

\section{General design criteria for a multiple species source}

Effusive ovens \cite{ross95} are simple and efficient atom sources.
They consist of an evacuated reservoir chamber in which the desired
species is stored in solid or liquid form, and is in equilibrium
with its vapor. The vapor effuses through a small opening towards
the experiment. The flux is easily controlled by changing the oven
temperature and thus the equilibrium vapor pressure in the
reservoir.

The relation between the flux and the vapor pressure is linear as
long as the vapor pressure is low enough that the flow through the
oven nozzle is molecular. Equivalently, the mean free path at the
vapor pressure is bigger than the nozzle size. Operation beyond the
linear regime results in viscous or supersonic flows, characterized
by higher total atom fluxes and by a depletion of the low velocity
tail of the Maxwell distribution. Zeeman slowing usually captures
the low-velocity tail and therefore requires operation of the oven
in the linear regime.

If multiple atomic or molecular species must be present in the beam,
loading a single reservoir with more than one pure substance is in
general not appropriate. The vapor pressures at the same temperature
are likely to be different. The ratio of fluxes is then fixed at
approximately the ratio of the vapor pressures, which has a weak
dependence on temperature. In the case of our experiment, sodium has
a vapor pressure three orders of magnitude higher than lithium at
the same temperature, meaning that at optimum sodium flux the
lithium flux is too low, and at the optimum lithium flux the sodium
consumption rate is impractical.

\begin{figure}
    \begin{center}
    \includegraphics [width=3.2in] {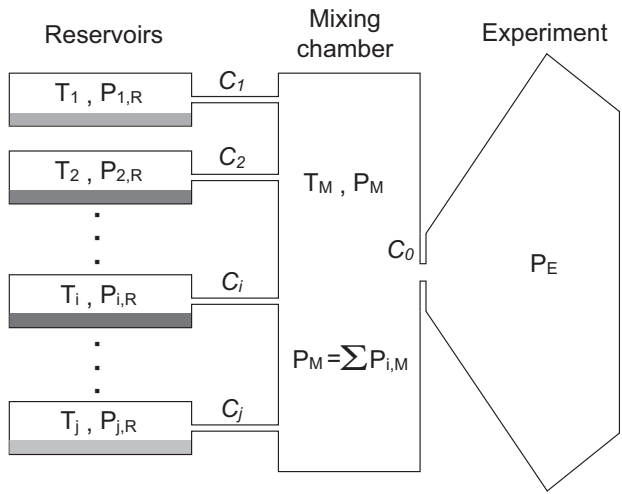}
    \caption[Title]{Design of a multiple species oven. Reservoirs
    containing pure components are connected to a mixing chamber through
    mixing nozzles. The mixing chamber is connected to the experiment
    through the oven nozzle. $P_{i,R}$ are the saturated vapor pressures
    at the reservoir temperatures $T_i$. In the mixing chamber a gaseous
    mixture at temperature $T_M$ is formed, with partial pressures
    $P_{i,M}$. $P_E$ is the pressure in the experiment chamber. Nozzle
    vacuum conductances are shown in italics. For proper operation,
    $P_{i,R}\gg P_{i,M}\gg P_E$.} \label{fig:multipleoven}
    \end{center}
\end{figure}

A solution to this problem is the use of multiple reservoirs, each
holding a pure component, connected to a mixing chamber
(Fig.~\ref{fig:multipleoven}). The design requirements for all
reservoirs is that they should deliver their component to the mixing
chamber at an adjustable rate. In steady state, the mixing chamber
should not absorb the components. Then, the flux of one species
through the oven nozzle equals the flux through the mixing nozzle,
and can be tuned by changing the temperature of the reservoir.

For simplicity we will assume that the pressures in the reservoirs
are much bigger than the pressure in the mixing chamber, which in
its turn is much bigger than the pressure outside the oven. Such
pressure ratios are also the best choice for oven operation, as it
will be seen below.

The throughput of the $i$th component, $Q_{i,R}$, from the $i$th
reservoir to the mixing chamber is given by
$$Q_{i,R}=C_i(P_{i,R}-P_{i,M})\simeq C_iP_{i,R}$$ where $Ci$ is the
conductance of the $i$th mixing nozzle, $P_{i,R}$ the vapor pressure
in the $i$th reservoir, and $P_{i,M}$ the partial pressure in the
mixing chamber. Assuming that the reservoirs are not contaminated
with different species, and that $P_{i,M}$ is smaller than the
saturated vapor pressure in the mixing chamber, the throughput of
the component from the oven to the experiment is given by
$$Q_{i,T}=C_0P_{i,M}=Q_{i,R}-P_{i,M}\sum_{j\neq i}C_j$$ where $C_0$
is the vacuum conductance of  the oven nozzle and $P_{i,M}$ is the
partial pressure of component $i$ in the mixing chamber. The second
term in the sum denotes the loss of component $i$ by backflow into
other reservoirs. It can be made negligible if all the mixing nozzle
conductances $C_j$ are much smaller than $C_0$.

In steady state operation, assuming that all mixing nozzles have
conductances much smaller than the oven nozzle,
$$Q_{i,T}=Q_{i,R}=C_iP_{i,R}$$ and the flux of component $i$ from
the oven can be tuned by heating the reservoir in the same way as
for a simple effusive oven.

Backflow of other species from the mixing chamber into the reservoir
is not desirable, for two reasons. First, these other species will
be lost into the reservoir rather than making the effusive beam.
Second, different species could react with or dissolve into the pure
substance loaded in the reservoir, reducing its vapor pressure
\cite{dush62}.

Reducing the backflow to negligible levels and even complete
suppression can be achieved by a proper design of the mixing
nozzles. Backflow occurs through diffusion against the constant
stream of atoms or molecules coming from the reservoir. If this
stream is sufficiently fast it will blow away diffusing components
before they could reach the reservoir.

We have estimated the parameters of the mixing nozzle by requiring
that in the nozzle the diffusion speed of component $j$ into
component $i$ is smaller than the macroscopic speed of the stream.
For a nozzle made of a long cylindrical tube, the approximate
condition for suppression is
$$\lambda_{j,P_{i,R}} < d$$ where $\lambda_{j,P_{i,R}}$ is the mean
free path of species $j$ in the gas of $i$ at reservoir $i$ pressure
$P_{i,R}$, and $d$ is the nozzle diameter. An equivalent statement
is that the flow in the mixing nozzle should be viscous or at least
in the intermediate regime.

Chemical reactions and solution formation is also a concern for the
mixing chamber, but negative effects can be diminished by raising
the temperature of the chamber. As long as the partial pressures of
different components are below their saturated vapor pressure,
condensation will not occur, and the rate of chemical reactions will
be significantly decreased. The guidelines for temperature settings
are completed by the condition that the nozzles should have the
highest temperatures in the oven to avoid clogging.

\section{The sodium-lithium atom source}

\begin{figure}
    \begin{center}
    \includegraphics [width=3.2in] {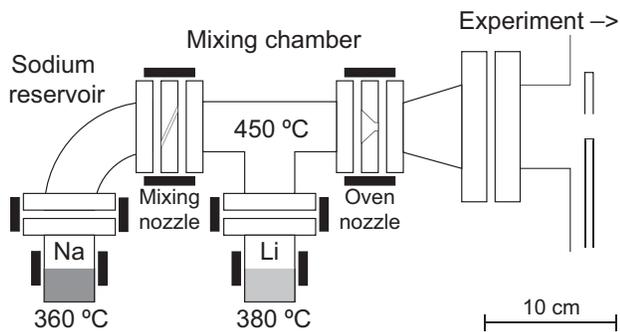}
    \caption[Title]{Exploded cross section view, to scale, of
the \na\ -\li\ source. The oven is built from 2.75 inch ConFlat
fittings, made from 316 SS. Alkali receptacles are capped half
nipples, and the nozzles are machined from double-sided blanks. The
2 mm diameter mixing nozzle is angled to achieve a length of 32 mm,
and the thin oven nozzle has a diameter of 4 mm. A copper plate with
a 7.5 mm hole, shown at the right edge of the figure, acts as a
collimation aperture. Black rectangles represent the six resistive
band heaters used for temperature adjustment. The shown temperatures
of the alkali receptacles and of the mixing chamber are typical for
double species operation.} \label{fig:nalioven}
    \end{center}
\end{figure}

A schematic of our \na\ -\li\ oven is given in
Fig.~\ref{fig:nalioven}. The oven follows the basic design
principles described above, but its construction was simplified by
putting the lithium in the mixing chamber rather than in its own
reservoir. This modification was necessary to keep the maximum oven
temperature at the specified operation limit of ConFlat knife-edge
vacuum flanges, 450 $^{\text{o}}$C. The vapor pressure required for
proper operation of a lithium reservoir must be tens of times bigger
than the vapor pressure in the mixing chamber, and can be achieved
only at higher temperatures.  Building the oven from ConFlat parts
makes it compatible with the rest of our vacuum chamber, and allows
easy assembly and disassembly during alkali reloading.

Since the lithium is placed in the mixing chamber, chemical reaction
and alloy formation are possible. The binary phase diagram of sodium
and lithium \cite{mass90} exhibits immiscibility regions. Above
lithium's melting temperature, 180.6 $^{\text{o}}$C, the mixture is
liquid at all concentrations, but it can phase separate into
sodium-rich and lithium-rich liquids. Above the critical
temperature, 303.2 $^{\text{o}}$C, the sodium and the lithium are
fully miscible.

A small amount of sodium always dissolves into the lithium during
two species operation. Above the critical temperature there is only
one liquid phase, a solution of sodium in lithium. The concentration
of sodium can be estimated by assuming that the solution is ideal
and thus obeys Raoult's law. Raoult's law states that the vapor
pressure of the solute is smaller than the saturated vapor pressure
by the atomic concentration of the solute in the solvent. The vapor
pressure of sodium is three orders of magnitude higher than
lithium's at the same temperature. In an alloy above which both
elements have equal partial vapor pressures, the atomic
concentration of sodium is then only 0.1\%. The sodium-lithium
mixture exhibits tendency towards phase separation. Therefore a
positive deviation from Raoult's law is expected \cite{lee99}, and
the real sodium concentration is smaller than our estimate. Since
the sodium concentration is small, lithium vapor pressure is not
decreased significantly by the presence of sodium.

The sodium-lithium oven is built from 2.75 inch ConFlat fittings.
The presence of lithium, which reacts with metals more
aggressively than other alkali, and the relatively high operating
temperatures required the use of non-standard materials for the
fittings and the gaskets. Copper gaskets are the standard choice,
but when heated to 400-450 $^{\text{o}}$C they can bond to the
knife edges, making disassembly difficult. Annealed nickel gaskets
were chosen for their high temperature performance and for their
good compatibility with molten alkali metals.

Nickel gaskets, even when fully annealed, are harder than copper
gaskets, and their properties vary among manufacturers and different
batches. As a result, reliable sealing and resealing was difficult
when we used standard ConFlat fittings made from type 304 stainless
steel. The knife edge dulls after only a few reseals, probably
because of annealing of type 304 when it is cycled from room
temperature to 400-450 $^{\text{o}}$C.

We have overcome this problem by using slightly modified fittings
made from type 316 stainless steel. A chromium-nickel steel, type
316's composition differs from type 304's by the addition of 2-3\%
percent molybdenum. It is offered as a material option by vacuum
manufacturers for its better high-temperature strength. We have
found that type 316 fittings were indeed superior to standard type
304 fittings between 400 and 450 $^{\text{o}}$C and could be
resealed multiple times.

The fittings were modified by removing a 0.127 mm layer of material
from the flange faces. By doing this the knife edge recession is
reduced and the knife can cut deeper into the nickel gaskets, which
we often found to be thinner than industry specifications. The
modified fittings were manufactured by A\&N Corporation \cite{a&n}.

The oven nozzle has a diameter of 4 mm, equal to the narrowest
collimating aperture, a differential pumping tube. The mixing nozzle
is a 2 mm diameter, 32 mm long tube angled with respect to the axis
of the machine. Both nozzles were machined in double-sided ConFlat
blanks.

For heating, band heaters are placed around flanges and around
alkali receptacles. We have used custom Mi-Plus heaters manufactured
by TEMPCO \cite{tempco}. They can be operated at temperatures up to
760 $^{\text{o}}$C, and have incorporated K-type thermocouples,
ensuring reproducible temperature readout. The heater powers are 100
W for the alkali receptacles, 150 W for the receptacle flanges, and
300 W for the nozzle flanges.

The heaters are controlled by commercial temperature controllers.
The oven is thermally insulated by wrapping first ceramic fiber tape
to form a 1 cm thick layer of insulation. Two or three layers of
household use aluminum foil are wrapped on top of this layer.
Temperature stability is approximately 0.1 $^{\text{o}}$C.

The oven is connected to a section of our apparatus in which the
pressure is $1\times10^{-8}$ Torr. In this section the beam is
collimated to $5\times10^{-5}$ sr divergence by a differential
pumping tube with an inner diameter of 4 mm.

\section{Operation and performance}

The oven is loaded with 25 g of \na\ and 10 g of isotopically
enriched \li\ (95\% purity). Sodium is commercially available in
sealed glass ampoules which we break just before loading. Enriched
\li\ is available as chunks under mineral oil, and has to be cleaned
prior to loading. We had the \li\ cleaned and repackaged into sealed
glass ampoules by a materials preparation laboratory \cite{mpc}.

The target alkali temperatures are achieved by setting the
receptacle heaters to these temperatures, 360 $^{\text{o}}$C for
sodium and 380 $^{\text{o}}$C for lithium. The heaters around the
flanges of the receptacles are set to 20 $^{\text{o}}$C above the
alkali temperature for a smooth thermal gradient from the alkali
receptacles to the nozzles. Nozzle heaters are both set to 450
$^{\text{o}}$C. The nozzles are the hottest areas in the oven to
prevent their clogging. When the experiment is not running, nozzle
heaters are kept on while the other heaters are turned off. From
this idle state, it takes less then 10 minutes to heat the oven to
its operating temperatures.

The mixing nozzle operates in the viscous flow regime. Its
conductance, 0.08 L/s, is difficult to calculate in this regime, and
was measured from the sodium consumption rate. At 450 $^{\text{o}}$C
the calculated molecular flow conductance of the oven nozzle is 2.6
L/s for sodium and 5 L/s for lithium. This results in a ratio of
sodium pressures in the reservoir and the mixing chamber of
approximately 30. The partial pressures in the mixing chamber are
$4\times10^{-3}$ Torr for sodium and $4\times10^{-5}$ Torr for
lithium.

Alkali-alkali collision cross sections at room temperature are quite
large, making the mean free paths in alkali vapor considerably
shorter than those encountered in normal vacuum practice. We have
estimated the cross sections using the Massey-Mohr formula
\cite{mass34}:
$$\sigma=5\times10^{11}(C_6/v)^{2/5}$$ where $\sigma$ is the
collision cross-section in cm$^2$, $C_6$ the Van der Waals
coefficient in erg$\cdot$cm$^6$, and $v$ the speed in cm/s. For
sodium we have used the experimental value of $C_6$ given in
\cite{buck65} and for lithium the theoretical calculation given in
\cite{font61}.

The estimated mean free path of lithium in the sodium reservoir is
0.1 mm, satisfying the condition for backflow suppression given in
the previous section. The estimated mean free paths for sodium and
lithium in the mixing chamber are 1.7 and 2.8 mm, slightly lower
than required for molecular flow through the 4 mm oven nozzle. In
this regime the oven nozzle conductance, calculated using
conductance tables given in \cite{dush62}, is not significantly
changed from its molecular flow value.

Total sodium atom flux, measured from the sodium consumption rate,
is $1.6\times10^{17}$ s$^{-1}$. A lithium flux of $3\times10^{15}$
s$^{-1}$ was estimated from the vapor pressure. The fluxes in the
collimated beam are approximately $10^{4}$ times lower. The limits
on flux tuning are given by different factors for sodium and
lithium. For sodium, the minimum operating temperature is
approximately 200 $^{\text{o}}$C, the equilibrium temperature with
only the nozzle heaters on. The upper limit, 360 $^{\text{o}}$C, is
given by the requirement that the flow through the oven nozzle is
molecular. For lithium the upper limit, 435 $^{\text{o}}$C, is given
by the requirement that the nozzles are the hottest part of the
oven. We have decided not to operate the lithium reservoir below the
critical temperature, 303.2 $^{\text{o}}$C, to avoid a possibly
complicated dependence between reservoir temperature and atom flux.
The atomic fluxes can be varied over three orders of magnitude,
between $1.3\times10^{14}$s$^{-1}$ and $1.6\times10^{17}$s$^{-1}$
for sodium, and from $5\times10^{13}$ s$^{-1}$ to $3\times10^{16}$
s$^{-1}$ for lithium.

The Zeeman slower originally designed for sodium was modified for
double operation by overlapping lithium laser slowing light with
the sodium slowing light. With this double atom source we can load
sodium, lithium, or overlapping double species MOTs. The lithium
MOT traps $3\times 10^8$ atoms and loads in approximately 4 s. The
sodium MOT traps $10^{10}$ atoms and loads in approximately 2 s.
The atom numbers given are accurate within a factor of 2. For the
double species MOT, the number of lithium atoms is reduced to
approximately half, as measured from the drop in MOT fluorescence
intensity. The number of sodium atoms is not changed.

The double species MOT provides an excellent starting point for
further cooling the mixture into quantum degeneracy. Bose-Einstein
condensates of \na\ with $2\times10^{7}$ atoms and degenerate \li\
Fermi gases with $5\times10^{7}$ atoms are routinely produced
\cite{hadz03big_fermi}.

With a full 25 g sodium load, the oven operates continuously for
1200-1300 hours. We have estimated that the 10 g lithium load should
last for 10000 hours at the maximum flux. Sodium changes are
performed twice a year. We do not find any lithium deposits in the
sodium reservoir, which proves that the mixing nozzle design
suppresses lithium backflow. The low conductance mixing nozzle
allows clean venting with high purity argon during the changes, and
baking is not needed. The experiment can be run again within hours
of sodium reloading.

Most of the effused sodium deposits on the collector plate and on
the walls of the 4.5 inch ConFlat 6-way cross to which the oven is
attached. Sodium needs to be cleaned every 2-3 years. The deposition
occurs in readily accessible areas and the cleaning procedure takes
at most a few days.

\section{Discussion}

Efficient loading of a multiple-species MOT using a multiple species
oven requires the use of a Zeeman slowing scheme which operates for
all species. While laser beams with different frequencies can be
easily overlapped, the solenoid coil which generates the spatially
varying magnetic field is the same for all species.

It is possible to design a solenoid which will operate for more than
one alkali atom. The most important parameter for the design of a
Zeeman slower is the maximum acceleration of an atom interacting
with a resonant laser beam:
$$a_{max}=\frac{\pi\hbar\Gamma}{m\lambda}$$
where $\Gamma$ and $\lambda$ are the linewidth and the wavelength of
the resonant light, and $m$ is the mass of the atom. In the case of
alkali atoms, it is the atom mass which varies most from species to
species. The maximum acceleration decreases from lithium to cesium.

The construction of the slowing coil defines the maximum speed of
the atoms which can be slowed (the capture velocity), and the
acceleration of the slowed atoms. This acceleration must be smaller
than $a_{max}$. Since $a_{max}$ decreases with mass, a coil designed
for a given alkali will work also for lighter species. However, at
the same capture velocity and same temperature a smaller part of the
Maxwell-Boltzmann distribution can be slowed for the lighter atom.
In our experiment, the performance of sodium-designed coils at
slowing lithium was satisfactory.

The use of single beam, multiple atom source can significantly
simplify the design of a multiple species laser-cooling experiment.
The general design criteria given in Sec. II is applicable for any
mixture of pure substances as long as chemical reactions do not
occur between the components. The trichamber thallium-sodium oven
described in \cite{rega02}, developed independently from ours, uses
the same principle as our sodium-lithium source.

The ConFlat construction we adopted for our sodium-lithium oven does
not require complicated machining and can be applied to many of the
atomic isotopes used in laser-cooling experiments. The main
limitation of the ConFlat design using 316SS parts and nickel
gaskets is its maximum operating temperature. Multiple reassembly
can be easily achieved at temperatures up to 400 $^{\text{o}}$C. We
have been able to reassemble a flange heated to 500 $^{\text{o}}$C
and we have learned that the seals are reliable to at least 650
$^{\text{o}}$C, although they might not be resealable.

Given the typical atomic fluxes needed in laser-cooling experiments,
partial pressures in the mixing chamber should be around
$1\times10^{-3}$ Torr. Alkali isotopes of K, Rb and Cs all have
vapor pressures higher than sodium. Ca, Sr, and Yb isotopes which
have been laser-cooled to high phase space densities
\cite{binn01,kato99,taka03} have vapor pressures comparable with
lithium. The design described here should work for all these
species.

\begin{acknowledgments}

This work was supported by the NSF, ONR, ARO, and NASA. We thank
Michele Saba and Zoran Hadzibabic for a critical reading of the
manuscript.

\end{acknowledgments}

\end{document}